\documentclass[12pt]{article}
\begin{document}
\title{Comment on relationship between atomic squeezed and entangled states}
\author{Arup Banerjee\\
Laser Physics Division, Centre for Advanced Technology\\
Indore 452013, India}
\date{}
\maketitle
\begin{abstract}
The ability of parameter recently proposed by Sorensen et al. (Nature ${\bf 409}$, 63 (2001)) to characterize entanglement or inseparability is investigated
for a system of two two-level atoms interacting with a single mode of radiation field. For comparison we employ the
necessary and sufficient criterion of Peres and Horodecki for inseparability of bipartite states. We show that for diagonal
atomic density matrix the parameter of Sorensen et al. fails to show entangled nature of collective atomic state whereas same
state is inseparable in accordance with Peres-Horodecki criterion.
\end{abstract}
\section*{}
Recently quantum entangled states have been focus of both theoretical and experimental research.
This is due to the realization that the quantum entanglement is a key resource for implementation
of protocols of quantum computation and quantum information processing \cite{physics}. Theoretically
lot of work has been done to quantify the quantum entanglement. For detail see the recent review by Horodecki et al. \cite{rev}. 
Simultaneously several schemes have been proposed for generation of these states in many 
physical systems. In connection to the generation of multi particle entanglement in Bose-Einstein
condensate Sorensen et al \cite{sorensen} proposed a parameter to characterize the entanglement. This parameter 
is given by
\begin{equation}
\xi^{2} = \frac{N\left (\Delta S_{\vec{n}_{1}}\right )^{2}}{\langle S_{\vec{n}_{2}}\rangle^{2} + 
\langle S_{\vec{n}_{3}}\rangle^{2}}
\end{equation}
where $S_{\vec{n}} = \vec{n} \cdot \vec{S}$, $\vec{S}$
is angular momentum operator representing the collective atomic operator
and $\vec{n}$'s are mutually orthogonal unit vectors. The value of the parameter going below unity, that is 
$\xi^{2} < 1$, signifies atomic entanglement or quantifies inseparability of collective atomic states. It
is important to note here that the same parameter is also used to characterize atomic squeezed 
states. Thus the collective atomic states satisfying $\xi^{2} < 1$ are also referred to as atomic
squeezed states. Moreover the parameter $\xi^{2}$ is also related to other parameters proposed
for characterizing atomic squeezed states.

In this paper our aim is to investigate the capability $\xi^{2}$ given by Eq.(1) to characterize the
entanglement properties of bipartite states. As is well known that for both pure or mixture bipartite 
states there exists a necessary and sufficient criterion for inseparability of composite state
due to Peres and Horodecki \cite{peres,horodecki}. According to this criterion separability
is signified by existence of non-negative eigenvalues of partial transpose of density matrix of the
composite system. For our purpose we study the entanglement property of a model system of two two-level atoms
interacting with a single cavity mode by using both Peres-Horodecki
criterion and the parameter defined in Eq.(1). We demonstrate that for a particular case of initial atomic and cavity
field states the collective atomic state at any later time is always inseparable in accordance with the Peres-Horodecki
criterion. On the other hand under the same initial condition $\xi^{2}$ is always greater than unity. In the following we 
present our analysis and also discuss the condition under which $\xi^{2}$ can characterize the entanglement property of
composite systems.

We consider two two-level atoms with ground and excited states denoted by $|g_{i}\rangle$ and $|e_{i}\rangle$ respectively,
where $i=1,2$, coupled to a single cavity radiation mode with same coupling constant $g$. 
The interaction hamiltonian for the two atom system with a single cavity mode is given by
\begin{equation}
H = g\sum_{i}\left (S_{i}^{+}a + S_{i}^{-}a^{\dagger}\right )
\end{equation}
where $a (a^{\dagger})$ is the annihilation (creation) operator of the radiation mode, $S_{i}^{+}=|e_{i}\rangle\langle g_{i}|$
and   $S_{i}^{-}=|g_{i}\rangle\langle e_{i}|$ represent the atomic raising and lowering operators respectively. Recently Kim
et al. \cite{kim} employed the same system to investigate the generation of the entangled state of the atomic system when the field
mode is in number and thermal states. Here we consider the case of both the atoms in their ground states ($|g_{1},g_{2}\rangle$)
and the field mode in the number state $|n\rangle$. The atomic density operator for the above mentioned initial condition at $t>0$
is given by
\begin{equation}
\rho(t) = X_{1}|1\rangle\langle 1| +  X_{2}|2\rangle\langle 2| + X_{3}|3\rangle\langle 3| 
\label{3}
\end{equation} 
where $|1\rangle = |e_{1},e_{2}\rangle, |2\rangle = \frac{1}{\sqrt{2}}\left (|e_{1},g_{2}\rangle + |g_{1},e_{2}\rangle\right 
)$ and $|3\rangle = |g_{1},g_{2}\rangle$ and the coefficients $X_{i}$ are given by 
\begin{eqnarray}
X_{1} & = & \frac{n\left (n - 1\right )}{(2n - 1)^{2}}\left [\cos^{2}\lambda t - 2\cos\lambda t + 1\right] \nonumber \\
X_{2} & = & \frac{n}{(2n - 1)}\sin^{2}\lambda t \nonumber \\
X_{3} & = & \frac{n^{2}\cos^{2}\lambda t + 2n(n -1)\cos\lambda t + (n - 1)^{2}}{(2n - 1)^{2}}
\label{4}
\end{eqnarray}
where $\lambda = gt\sqrt{2(2n - 1)}$. In accordance with the Peres-Horodecki criterion the collective atomic states at $t>0$ are 
entangled or inseparable atomic if and only the condition $X_{2}>\sqrt{X_{1}X_{3}}$ is satisfied. For the single photon case,
that is $n=1$, this condition reduces to $\sin^{2}2gt > 0$. Therefore for fock state $|1\rangle$ and both the atoms in their ground
state collective atomic state is always entangled except at $t=0$ \cite{kim}. On the other hand for the same initial atomic
and cavity mode states the squeezing parameter of Eq.(1) is given by
\begin{equation}
\xi^{2} = \frac{1 + \sin^{2}2gt}{\cos^{4}2gt}.
\label{4}
\end{equation}
It is easy to verify that $\xi^{2}$ given above (Eq.(\ref{4})) never goes below unity at any time signifying that collective
atomic states are not entangled. This also shows that no atomic squeezed states are generated in the atom-field interaction.
Thus we show that a density matrix of composite system of $2\times 2 $ dimension can be inseparable or entangled in 
accordance with the Peres-Horodecki
criterion. However, same system fails to satisfy the condition of entanglement given by Eq.(1).  

We emphasize here that the condition $\xi^{2}<1$ to signify entanglement or atomic squeezed state require that the density
matrix of composite must have non-zero off-diagonal ( coherence ) matrix elements \cite{agarwal,banerjee}. To show this 
explicitly we consider the density matrix of the form
\begin{eqnarray}
\rho & = & X_{1}|1\rangle\langle 1| +  X_{2}|2\rangle\langle 2| + X_{3}|3\rangle\langle 3| \nonumber \\
     & +  & Y|1\rangle\langle 3| + Y|1\rangle\langle 3|.
\label{5}
\end{eqnarray} 
For the sake of simplicity we assume $Y$ to be real. It is straightforward to extend the discussion for a complex $Y$ too. Now 
corresponding to the above density matrix (Eq.(\ref{5})) the parameter $\xi^{2}$ is given by
\begin{equation}
\xi^{2} = \frac{2Y + 2 - \langle S_{z}^{2}\rangle}{\langle S_{z}\rangle^{2}}
\end{equation}
where $S_{z} = \frac{1}{2}\left ( S^{+}S^{-} - S^{-}S^{+}\right )$. The entanglement or squeezing  condition requires that
\begin{equation}
\langle S_{z}^{2}\rangle + \langle S_{z}\rangle^{2} > 2 + 2Y
\label{6}
\end{equation}
For $Y=0$, that is for a diagonal density matrix Eq.(\ref{6}) can never be satisfied as $\langle S_{z}^{2}\rangle$,
$\langle S_{z}\rangle^{2} \leq 1$. However for, $Y \neq 0$ it is still possible to satisfy Eq.(\ref{6}) under some
conditions. For example, if the field is in a coherent or a squeezed state then off-diagonal terms of the atomic 
density matrix are not zero and it may be possible to achieve $\xi^{2} < 1$ \cite{wineland,ueda} under appropriate condition.  
 
In conclusion we have demonstrated that the parameter proposed in Ref.\cite{sorensen} is not suitable for 
characterizing entanglement or inseparability properties of composite systems having diagonal density matrix. It is only for systems with
off-diagonal density matrix $\xi^{2}$ may be employed to characterize entangled states.
 

\end{document}